\documentclass[aps,showpacs]{revtex4}

\usepackage{graphicx}

\newcommand{\eqref}[1]{(\ref{#1})}

\renewcommand{\d}{\partial}

\newcommand{\nn}{\nonumber\\}

\newcommand{\rh}{\varrho}

\newcommand{\ep}{\varepsilon}

\renewcommand{\k}{{\bf k}}
\newcommand{\p}{{\bf p}}

\newcommand{\Disc}{\mathop{\textrm{Disc}}}
\renewcommand{\Im}{\,\textrm{Im}\,}
\renewcommand{\Re}{\,\textrm{Re}\,}
\newcommand{\pint}[2]{{\int\!\frac{d^{#1}#2}{(2\pi)^{#1}}\,}}
\newcommand{\pintz}[1]{{\int\!\frac{d #1}{2\pi}\,}}

\newcommand{\MSbar}{{\ensuremath{\overline{\mathrm{MS}}}}}

\newcommand{\sgn}{{\ensuremath{\mathop{\mathrm{sgn}}}}}

\renewcommand{\c}[1]{{\cal{#1}}}

\newlength{\szovszel}\newlength{\szovmag}

\begin{document}

\title{Renormalization of 2PI resummation: a renormalization scheme approach}
\author{Antal Jakov\'ac} 
\email{Antal.Jakovac@cern.ch}
\affiliation{Institute of Physics, Budapest University of Technology
  and Economics, Budafoki \'ut 8, H-1111 Budapest , Hungary}
\date{\today}

\begin{abstract}
  A practical method is suggested for performing renormalized 2PI resummation
  at finite temperature using specific momentum dependent renormalization
  schemes. In this method there is no need to solve Bethe-Salpeter equations
  for 2PI resummation. We examine the consistency of such schemes in the
  paper. The proposed method is used to perform a two-loop renormalized 2PI
  resummation in the finite temperature $\Phi^4$ model.
\end{abstract}


\maketitle

\section{Introduction}

By using direct perturbation theory (ie. strict power series of some coupling
constant) to calculate finite temperature or nonequilibrium correlation
functions, one very frequently encounters infrared (IR) sensitive
contributions, ie. contributions that become very large (occasionally
infinite) when we approach a certain energy scale. These contributions, if
they show up consequently in every higher loop level, render the direct
perturbation theory non-convergent. Then we have to collect the problematic
diagrams and sum them up, before the actual perturbative procedure begins.
Since these diagrams are usually also ultraviolet (UV) divergent besides their
IR sensitivity, the resummation of these diagrams must be accompanied by a
proper renormalization.

In the literature there are several approaches to the renormalization of
resummed perturbative series. One of them treats the resummation in the
original sense, as sum of diagrams belonging to a certain subset of all
Feynman diagrams. One possibility is to apply BPHZ renormalization: then,
without the reference to the Lagrangian we can assign a finite (renormalized)
value to every Feynman diagram \cite{Collins}, and we define the renormalized
value of the sum of diagrams as sum of their BPHZ renormalized value. In
principle this method can be applied to all kind of resummations, but
technically it can be rather involved. The difficulty is to be able to work
out the BPHZ forest for an arbitrary diagram in the relevant subset. This task
was performed for the 2PI resummation in various models in
\cite{HeesKnoll}-\cite{2PIgauge} by solving a set of Bethe-Salpeter equations
for the four point functions. Diagrammatic analysis can be used also to work
out counterterms: such a technique was applied in \cite{Verschelde} to
renormalize the 2PPI resummation. Diagram by diagram renormalization can be
applied for large N bubble resummation \cite{PSz}, too.

Another approach is based on the idea that we can try to modify the generating
equation of the resummation in order to obtain finite results. It is not
guaranteed to work, since the equation that we use is, in general, not exact,
and so we may miss some terms important for UV finiteness. In the
Schwinger-Dyson resummation it seems to be an important obstacle, cf.
\cite{Cooperetal}. On the other hand, for Hartee-Fock resummation, such
improvement exists, cf. \cite{Destrietal}. In the O(N) model the resummation
is governed by the $1/N$ order of the diagram. Here we can write up closed
forms, at least in the lowest orders, and so one expects that it can be also
renormalized in closed form. In \cite{ON1perN} the authors present
renormalization of the O(N) model up to $\c O(1/N)$ order, at the physical
point.

A third approach uses the fact that the contribution of a diagram depends on
the renormalization scheme we use. It may be possible to find a scheme where
the IR sensitive part of all diagrams disappears. In this IR safe scheme there
is no need for resummation, usual perturbation theory can be used to calculate
physical observables. Thereafter we can change the renormalization scheme to a
standard one (eg. \MSbar) by appropriate change of the renormalized parameters
of the Lagrangian (``renormalization group'' transformation). At the end of
this procedure we obtain the same result in the \MSbar\ and the IR safe scheme
for all quantities up the order $(\mathrm{coupling})^n$ where $n$ is the order
of calculation. The difference is the resummation effect: it is responsible
for the difference that one scheme is IR safe while the other is not.

These thoughts was exploited in \cite{JakovacSzep} to perform a renormalized
momentum independent resummation as a change of renormalization scheme from an
IR safe one to \MSbar. The IR safe scheme was constructed in a way that if in
a perturbative calculation we encounter IR sensitive contributions, then we
choose the finite part of the counterterm to cancel this IR sensitive part.
This guarantees in a natural way that the calculation is free from IR
sensitive contributions. At the end of the calculation we switch back to
\MSbar\ scheme with renormalization scheme transformation, thus we obtain the
final result in \MSbar\ scheme.

This idea is motivated by the ``environmentally friendly'' renormalization of
Ref. \cite{envfriend}, where the renormalization scale was used to take into
account the environmental dependence.

In \cite{AJ} the first step was done towards the generalization of the method
to momentum dependent resummations. It was observed, that if we do
perturbation theory with a propagator, which equals to the \MSbar\ propagator
\emph{asymptotically}, then all the overall divergences -- and so all the
counterterms -- are the same as in the \MSbar\ scheme. Since IR sensitiveness
occurs -- by definition -- in the IR, we expect that it can be captured by an
appropriate modification in the IR region. So if we base our perturbation
theory on a propagator that is exact in the IR and \MSbar-like in the UV, we
can use the \MSbar\ counterterms for renormalization. A great advantage of
this method is that one does not need to solve Bethe-Salpeter equations, which
is a central task in other, diagram-based 2PI renormalizations. On the other
hand it is not easy to implement the momentum independent resummation in this
method.

The purpose of the present paper is to connect the momentum independent and
momentum dependent method in a common framework. We will examine the
possibility and consistency of a perturbation theory with arbitrarily chosen
free Lagrangian, which not necessarily approaches the usual free part
($\frac12(\d\Phi)^2-\frac{m^2}2\Phi^2$) asymptotically. Therefore the
generated divergences are also different from the usual case. Still it can be
consistent, if the usual process of the renormalization can be accomplished --
the conditions for this will be examined in the paper. But once we have a
consistent perturbation theory with ``any'' propagator, then nothing prohibits
us to apply a scheme, where the renormalized self energy is \emph{zero}. In
this scheme the free propagator is exact -- and this is exactly the effect of
2PI resummation. So we can perform renormalized 2PI resummation with help of a
cleverly chosen renormalization scheme.

In this paper we accomplish this strategy for the two-loop 2PI resummation of
the $\Phi^4$ theory at finite temperature. The paper will be organized as
follows. In Section \ref{sec:PT} we study in detail the feasibility of the
renormalized perturbation theory with an arbitrary propagator, and work out
the strategy for 2PI resummation. In Section \ref{sec:Phi4} we perform
renormalized 2PI resummation in the quartic scalar model at finite temperature
up to two loop level. We close the discussion with Conclusions in Section
\ref{sec:conclusion}.

\section{Perturbation theory with non-conventional propagators}
\label{sec:PT}

Let us start from a renormalizable Lagrangian, which now means that it is
renormalizable in the non-perturbative sense: we can define regularization
dependent parameters in the Lagrangian in a way that the exact $n$-point
functions are all finite when we remove the regularization. We assume that we
split this bare Lagrangian into two pieces: a ``free'' part and an
``interaction part''. The free part should be quadratic in the fields in order
to be solvable, but there are also quadratic pieces in the interaction
Lagrangian (like the mass counterterms). Here, however, we do not fix the
kernel of the quadratic part to the form appearing in the bare Lagrangian: we
allow arbitrary choice of the kernel. We subdivide the interaction part, as
usual, assigning ``orders'' to them according to a formal loop parameter. In
this way we defined a perturbation theory. The question is: when is the
so-defined perturbation theory finite \emph{order by order}?  The counterterms
should be defined as the overall divergence of the 1PI diagrams, when
``overall divergence'' means that all the momenta at all lines are going to
the infinity at the same rate. When are these counterterms enough to make
finite the complete contribution of all the diagrams at a given order of
perturbation theory?

For a consistent renormalization, in fact, we should ensure only two
conditions \cite{Collins}. First, the divergence of an overall divergent
diagram should be local: in Fourier space it should be a polynomial of its
external momenta. Second, if we subtract all subdivergences of a diagram, the
diagram should be overall divergent or finite. The proof that these conditions
are satisfied in case of using the usual free propagator was done by Weinberg
\cite{Weinberg_thm}. But the basic property of the propagator, which was in
fact used by the proof, is that the free propagator can be power expanded
``around infinity'', ie.:
\begin{equation}
  \label{W_reg}
  G(p+k) = G(p) + k^\mu \d_\mu G(p) + \frac12 k^\mu k^\nu \d_\mu\d_\nu G(p) +
  \dots,
\end{equation}
if $p^\mu\to\infty$ (which means that all of its components go to infinity
with a fixed ratio). This ensures that differentiation with respect to
external momenta will decrease the degree of divergence of the diagram, and so
-- if we differentiate enough times -- the contribution will be finite. The
asymptotic expansion property of eq. \eqref{W_reg}, however, is true for a
wide class of functions, not just for $1/(p^2-m^2)$. With any of these
functions the Weinberg theorem goes through. Then all of the consequences will
hold true, too, which can be summarized in a phrase that the theory can be
renormalized by local counterterms with polynomial momentum dependence.

The Weinberg theorem in itself does not guarantee that we need a finite number
of counterterms, only that the counterterm method works. We also need the
condition that only those operators are divergent that show up in the bare
Lagrangian. In the four dimensional $\Phi^6$ model, for example, we generate a
logarithmic divergence for the 8-point function at one loop level, which was
not included in the original Lagrangian, and so it is not renormalizable. That
means that we should accept only those propagators, where there are no more
divergent diagrams than in the usual $1/(p^2-m^2)$ case. If we assume that the
propagator is asymptotically $G(p)\sim p^{-\alpha}$, then the one-loop
correction to the 6-point function in the $\Phi^4$ model contains the integral
$\sim \int d^4 p\,p^{-3\alpha}$. This is UV convergent if $\alpha>4/3$: so the
propagator should be smaller than $p^{-4/3}$ for asymptotic momenta.

The exact propagator in four dimensions picks up only logarithmic corrections
to the $1/k^2$ behavior, according to the Weinberg theorem. Thus it may be a
good candidate to be the basic propagator of a perturbation theory.

Once we have fixed the kernel, and so the propagator, too, we can use it for
ordinary perturbation theory. In particular, we can do finite temperature
calculations. Since temperature effects are exponentially suppressed in the
propagator at high momenta, the overall divergence of a diagram (when all the
momenta goes in the same rate to infinity) is \emph{temperature independent}.
This ensures that the counterterms are temperature independent, and so a zero
temperature renormalization is enough for a finite temperature calculation,
too.

According to the above analysis, a large class of functions is good to be used
as ``free'' propagators. But which one is worth to be used? The best choice is
the one which is optimized for the problem at hand, ie. where the perturbative
series converge the fastest. If we keep the bare Lagrangian intact, then any
choice describes the same physics, so each is a resummed version of another.
This is similar when we choose the best renormalization scale. To compute the
potential, for example, the best choice is to have a scale $\mu\sim 1/r$ where
$r$ is the distance between the point charges. Then large logarithmic
corrections are suppressed, perturbation series converge well. To keep the
bare Lagrangian intact we should tune the parameters of the perturbation
theory accordingly: in particular we have to work with the $\mu$-dependent
coupling $g(1/r)$. We will use the same strategy: we optimize the kernel for a
certain problem, for example for computations at temperature $T_0$. To keep
the bare Lagrangian intact we should work with $T_0$ dependent parameters in
the perturbation theory ($g(T_0)$). Then we describe the same physics from
different points of view.

Next we express this strategy in more formal terms in the $\Phi^4$ model.
First we establish the generic strategy, then we examine the relation of
perturbation theories with different kernel choices to each other.

\subsection{Generic strategy}
\label{sec:gen}

The bare Lagrangian of the theory reads
\begin{equation}
  \c L = \frac12 (\d\Phi_\mathrm{bare})^2 -
  \frac{m_\mathrm{bare}^2}2\Phi_\mathrm{bare}^2 -
  \frac{\lambda_\mathrm{bare}}{24} \Phi_\mathrm{bare}^4.
\end{equation}
In the usual procedure we first change to the renormalized field:
$\Phi_\mathrm{bare}^2=Z\Phi^2$, then we find
\begin{equation}
  \label{Lbare1}
  \c L = \frac12 (\d\Phi)^2 - \frac{m_0^2}2\Phi^2 -
  \frac{\lambda_0}{24} \Phi^4 +\frac{\delta Z}2(\d\Phi)^2,
\end{equation}
where $\delta Z=1-Z$, $m_0^2 = Zm_\mathrm{bare}^2$ and $\lambda_0=
Z^2\lambda_\mathrm{bare}$. Then we split $m_0^2 = m_R^2 +\delta m^2$ and
$\lambda_0=\lambda_R+\delta\lambda$ and find
\begin{equation}
  \c L = \frac12 (\d\Phi)^2 - \frac{m_R^2}2\Phi^2 -
  \frac{\lambda_R}{24} \Phi^4 +\frac{\delta Z}2(\d\Phi)^2  -
  \frac{\delta m^2}2\Phi^2 - \frac{\delta\lambda}{24} \Phi^4.
\end{equation}
The above form suggests a separation of the non-interacting and interacting
theory as
\begin{equation}
  \c L_0 = \frac12 (\d\Phi)^2 - \frac{m_R^2}2\Phi^2,\qquad \c
  L_I=-\frac{\lambda_R}{24} \Phi^4 +\frac{\delta Z}2(\d\Phi)^2
  -\frac{\delta m^2}2\Phi^2 - \frac{\delta\lambda}{24} \Phi^4,
\end{equation}
where we treat $\c L_I$ as interactions. It must be emphasized that the values
of the counterterms are not fixed beforehand: they are determined order-by
order, and they do depend on the choice of the chosen perturbative scheme. We
can prove \cite{Collins}, however, that in the $n$-th order of perturbation
theory the values of the counterterms are scheme independent up to $\c
O(\lambda^n)$.

The above described procedure is not unique. Starting from \eqref{Lbare1} we
can separate the renormalized and counterterms as
\begin{equation}
  {\cal L} = \frac12 \Phi\,K(i\d)\Phi -\frac\lambda{24}\Phi^4
  +\frac12 \Phi\,\delta K(i\d)\Phi -\frac{\delta \lambda}{24}\Phi^4.
\end{equation}
The kernel of the quadratic part (shortly ``kernel'' in the followings) used
here is arbitrary, although for consistency we will require certain properties
to be satisfied. To obtain the bare Lagrangian \eqref{Lbare1}, the kernel- and
the coupling-counterterms should satisfy
\begin{equation}
  \label{bare_ren_relation}
  Z p^2 - m^2_0 = K(p) +  \delta K(p), \qquad
  \lambda_0= \lambda_R + \delta \lambda.
\end{equation}
The above form suggests the following free - interaction separation
\begin{equation}
  \label{newren}
  \c L_0 = \frac12\Phi\,K \Phi,\qquad  \c L_I=
  -\frac{\lambda_R}{24} \Phi^4 + \frac12\Phi\,\delta K\Phi - \frac{\delta
    \lambda}{24}\Phi^4.
\end{equation}
The coupling constant and kernel counterterms are considered to be of higher
loop order as compared to the coupling constant and kernel, respectively.
Their values should be determined order by order, just like in the usual case.
We should compute the diagrams, determine the infinite parts and then make
them disappear by a suitable choice of the infinite part of the counterterms.
The propagator should satisfy \eqref{W_reg}, then the divergences are
polynomial in their momenta; it also has to have appropriate asymptotics, then
only those operators are divergent that are present also in the bare
Lagrangian. To make these divergences disappear, we should choose $\delta
K_\mathrm{div} = A p^2 - B$ and $\delta \lambda = C$, where $A,\,B$ and $C$
are divergent constants.  Since from \eqref{bare_ren_relation} we can write
$\delta K = \delta Zp^2 - \delta m^2$, so we can identify the infinite parts
of $\delta Z$ and $\delta m^2$ as $A$ and $B$, respectively. It must be
emphasized, that the actual form of $\delta Z, \delta m^2$ and $\delta
\lambda$ depends on the choice of the kernel. The logics of their
determination is the same as in the usual case when we use the kernel
$p^2-m^2$.

\subsection{Theories with different kernels}

The best strategy for perturbation theory would be to use that kernel that is
the most appropriate for the description of the given observable. The
background of working with different kernels is very similar to using
different schemes in the usual perturbation theory. If the bare Lagrangian is
the same, they should describe the same physics: now this is ensured by eq.
\eqref{bare_ren_relation}. Here the momentum dependence of $K$ and $\delta K$
should cancel each other, while $Z$ and $m_0^2$ are determined by the local,
divergent part of the kernel counterterm. Similarly, the coupling constant
counterterms is determined by the divergences of the four point function.

Although we determine the counterterms order by order, the following statement
ensures that in an $n$th order calculation the bare parameters of the
Lagrangian, as well as any observables are independent on the choice of the
kernel, up to $\c O(\lambda^n)$. As a consequence, at infinite order, we
recover the same bare Lagrangian, and obtain the same value for all
observables, with any choice of the kernel. Although the proof is essentially
the same as in the standard case (cf.  \cite{Collins}, in particular the
Chapter ``Oversubtractions'') it is worth to state here for our somewhat more
generic case.

\begin{list}{}{}
\item[\textbf{Statement}] \emph{In any regularized perturbation theory with
    kernel $K$ and kernel counterterm $\delta K$ which satisfy
    \eqref{bare_ren_relation} with $m_0^2 = m_R^2 + \delta m^2,\, Z=1+\delta
    Z$ and $\lambda_0=\lambda_R+\delta \lambda$ where $m_R^2$ and $\lambda_R$
    are fixed, and $\delta m^2,\,\delta Z$ and $\delta \lambda$ are determined
    order by order, the following statements are true in $n$-th order of the
    calculation:
  \begin{itemize}
  \item $\delta m^2,\,\delta Z$ and $\delta \lambda$ are independent on the
    kernel choice up to order $\lambda^n$
  \item any Greens function is kernel independent up to order $\lambda^n$.
  \end{itemize}
}
\end{list}

\begin{list}{}{}
\item[\textbf{proof}] First we observe that if we have a well defined
  renormalized theory with $K$ and $\delta K$, then in a theory with
  $K'=K-\overline{\delta K}$ and $\delta K'=\delta K + \overline{\delta K}$
  the bare Lagrangian is the same, and so, nonperturbatively, we obtain the
  same results for all observables. In this case \eqref{bare_ren_relation} is
  automatically satisfied for $K'$ and $\delta K'$ if it was true for $K$ and
  $\delta K$. This proves that in an infinite order perturbation theory the
  result is independent on the actual choice of the kernel. In finite order
  perturbation theory, however, order-by-order finiteness requires different
  counterterm structure for each kernels. We should prove that the results are
  stabilized below the order of the calculation $\c O(\lambda^n)$.

  At tree level there is no need for counterterms ($\delta K=0,\,\delta
  \lambda=0$), and \eqref{bare_ren_relation} ensures that $K = p^2-m_R^2\equiv
  K_0$.

  At one loop level we have a one-loop self-energy $\Sigma_1[K]$, which is
  only overall divergent. When the regularization is fixed then we can
  determine the ``infinite part'' of a diagram, so we write
  \begin{equation}
    \Sigma_1[K] = \Sigma^\mathrm{div}_1[K] + \Sigma^\mathrm{fin}_1[K].
  \end{equation}
  In order to ensure finiteness, we should choose the one loop kernel
  counterterm as $\delta K_1= \Sigma^\mathrm{div}_1[K] + \overline{\delta
    K}_1$, where $\overline{\delta K}_1$ is finite (but, possibly, momentum
  dependent). Then \eqref{bare_ren_relation} tells us that that the one-loop
  kernel $K_1$ has to satisfy
  \begin{equation}
    K_1+ \Sigma^\mathrm{div}_1[K_1] + \overline{\delta K}_1 = K_0 + \delta Z_1
    p^2 - \delta m_1^2.
  \end{equation}
  We then identify $\delta Z_1 p^2 -\delta m_1^2 = \Sigma^\mathrm{div}_1[K]$
  and $K_1=K_0 - \overline{\delta K}_1$. This latter equation states that
  $K_1=K_0+\c O(\lambda)$, therefore $\delta Z_1 p^2 -\delta m_1^2 =
  \Sigma^\mathrm{div}_1[K_0] +\c O(\lambda^2)$: this means that up to order
  $\lambda$ it is independent of the choice of $\overline{\delta K}_1$, ie. of
  the choice of the kernel.

  What happens at higher order? We write $\delta K_1 =
  \Sigma^\mathrm{div}_1[K] + \overline{\delta K}_1$. This means that we have a
  new quadratic counterterm to the Lagrangian with kernel $\overline{\delta
    K}_1$. At higher orders this counterterm should be included in any diagram
  $\Gamma[K]$.  Technically this means that we subsequently replace some
  propagators by $iG\to iG\cdot i \overline{\delta K}_1 \cdot iG$.  Since
  $G=K^{-1}$, this replacement means
  \begin{equation}
    iG\to \frac{\d iG}{\d K} \overline{\delta K}_1,
  \end{equation}
  in functional manner. So, up to $n$ inclusions into the diagram, it can be
  formulated as
  \begin{equation}
    \sum\limits_{\ell=0}^n \frac1{\ell!}\frac{\d^\ell \Gamma}{\d
      K^\ell}(\overline{\delta K}_1)^\ell.
  \end{equation}
  The zeroth term is $\Gamma$ itself, the $1/\ell!$ is the remnant of the
  $1/n!$ factor of the $n$th order in the perturbative series. We can write it
  as an infinite sum minus the difference. But the infinite sum is just the
  Taylor-series of $\Gamma[K+\bar \delta K_1] = \Gamma[K_0]$. So we find:
  \begin{equation}
    \sum\limits_{\ell=0}^\infty \frac1{\ell!} \frac{\d^\ell \Gamma}{\d
      K^\ell}(\overline{\delta K}_1)^\ell - \sum\limits_{\ell=n+1}^\infty
    \frac1{\ell!}\frac{\d^\ell \Gamma}{\d K^\ell}(\overline{\delta K}_1)^\ell
    = \Gamma[K+\overline{\delta K}_1] (1- \c O(\lambda^{n+1})) =  \Gamma[K_0]
    (1- \c  O(\lambda^{n+1})).
  \end{equation}
  We see, therefore, that the generated finite counterterm restores the value
  $\Gamma[K_0]$ up to order $\lambda^n$. This means that any renormalized
  observable, which is represented by the sum of diagrams $\Gamma$, is the
  same (kernel-independent) up to $\c O(\lambda^n)$. Since we did not use any
  specific property of $\bar\delta K_1$ which would exploit its one loop
  nature, the above statement can directly be generalized to higher orders.
  Moreover, the same line of thought can be followed for the coupling
  constant, when it is momentum independent. This completes the proof of the
  second part of the statement.

  The same proof also applies to the counterterm structure. Then we should use
  $\Gamma\to\Sigma^\mathrm{div}_1[K]$. At one loop level the value of
  $\Sigma^\mathrm{div}_1[K]$ defines the kernel counterterms. At higher
  ($n$th) loop level the inclusion of $\overline{\delta K}_1$ finite
  counterterm will generate divergent contribution, which has to be made
  vanish by the $n$th order kernel counterterm. So $\delta K_n$ contains the
  following contribution
  \begin{equation}
    \delta K_n \to \frac1{n!}\frac{\d^n \Sigma^\mathrm{div}_1[K]}{\d
      K^n}(\overline{\delta K}_1)^n.
  \end{equation}
  In the sum $\sum_{i=1}^n \delta K_i$, therefore, we recover the subseries:
  \begin{equation}
     \sum\limits_{i=0}^n \frac1{i!}\frac{\d^i \Sigma^\mathrm{div}_1[K]}{\d
       K^i}(\overline{\delta K}_1)^i = \Sigma^\mathrm{div}_1[K_0] + \c
     O(\lambda^{n+1}).
  \end{equation}
  So, although the one loop counterterm depends on the kernel, the higher
  order counterterm contributions coming from $\overline{\delta K}_1$
  inclusions ensure kernel independence up to $\c O(\lambda^n)$. Since this
  line of thought did not depend that we start at one loop level, it is valid
  for any other finite counterterm, and also the same line of thought applies
  to the coupling constant renormalization. This completes the proof of the
  first part of the statement.

  \emph{QED}
\end{list}

In practice, the most important question is the comparability of two
perturbation theories, without a reference to the bare theory. The above
statement claims that this can be ensured if we choose the parameters of
scheme ``$2$'' as $m_2(m_1,Z_1,\lambda_1), \,Z_2(m_1,Z_1,\lambda_1),\,
\lambda_2(m_1,Z_1,\lambda_1)$, where $m_1,\,Z_1$ and $\lambda_1$ are the
parameters of scheme ``$1$''. These relations can be found either by the
requirement that the bare Lagrangian is the same (first part of the above
Statement), or requiring that some observables has the same value in the two
schemes. This latter strategy is the \emph{matching}, which is a widely used
technique in field theory (cf. large mass decoupling
\cite{AppelquistCarrazone} or dimensional reduction \cite{dimred}). It works
in our case in a way that we compute three observables from theory ``$1$'',
and find those $m_2,\,Z_2,\,\lambda_2$ parameters that give the same result
for these observables from a calculation in theory ``$2$''. The above
statement then ensures that calculating any other observables, the difference
between the results of the two schemes is of $\c O(\lambda^{n+1})$ where $n$
is the order of the computation. The best is of course to choose that type of
observables which can be calculated quite reliably from both schemes of
perturbation theory.

The area where we want to apply the perturbation theory with modified kernel
is to optimize perturbation theory for finite temperature or in
non-equilibrium calculations. At each temperatures we can use a kernel, which
is optimized to that temperature. To compare the results at different
temperatures, however, we have to match the renormalized parameters. A good
set of quantities is the asymptotically large momentum regime observables at
$T=0$, since we expect that these observables should be independent of the
details of the kernel at finite momenta, but still sensitive to the values of
the renormalized parameters. In practical numerical calculations, of course,
we should choose quantities at large, but not asymptotic momentum, and
determine the matching procedure there.

\subsection{2PI resummation}

For optimizing the kernel for a given environment, we can use the condition
that the free propagator (the inverse of the kernel) is the exact one. Then at
any higher order diagrams the self-energy insertions, which would modify the
free -- in this case the exact -- propagator, should be missing. So we should
choose the kernel counterterms $\delta K$ at each perturbative order in a way
that it exactly cancels the corresponding self-energy $\Sigma[K](p)$,
calculated with the kernel $K$. On the other hand the diagrams with
self-energy insertions are exactly the 2-particle reducible diagrams.  These
are missing in this perturbation theory: therefore we realized 2PI
resummation, differently from the technique of Cornwall, Jackiw and Tomboulis
\cite{CJT}.

In formula, the total self-energy can be written, at $n$th order, as
\begin{equation}
   \Sigma^{(n)}_\mathrm{tot}[K](p) = \Sigma^{(n)}[K](p) - \delta K^{(n)}(p),
\end{equation}
where here (and in the followings) $\Sigma[K](p)$ means the self energy with
subdivergences subtracted. So 2PI resummation requires
$\Sigma^{(n)}_\mathrm{tot}[K](p)=0$ for all $n$: in this case the exact
propagator $G_\mathrm{ex}^{-1}(p) = K(p) - \Sigma_\mathrm{tot}[K](p) =
K(p)$ which is the inverse free propagator. So we should choose
\begin{equation}
  \label{2PIeq}
  \delta K^{(n)}(p) = \Sigma^{(n)}[K](p).
\end{equation}
Equation \eqref{bare_ren_relation} implies
\begin{equation}
  \label{gapeq}
  Zp^2 - m_0^2 = K(p) + \Sigma[K](p)
\end{equation}
at each perturbative order. It seems to be the usual 2PI equation, but it is
not quite so. In this formula the infinite part of $Z$ and $m_0^2$ is chosen
in a way that it cancels the infinities of $\Sigma[K](p)$: therefore,
implicitly, $Z$ and $m_0$ are functionals of $K$. By our Statement of the
previous subsection we can claim that at infinite order this dependence
vanishes; but at any finite order it is there.

To see a finite equation we use the fact that $\Sigma[K](p) =
A_\mathrm{div}p^2 -B_\mathrm{div} + \Sigma_\mathrm{fin}[K](p)$, since
$\Sigma$, by definition, is just overall divergent (subdivergences are
subtracted). Then we choose $Z= 1 + A_\mathrm{div} + \delta\zeta$ and
$m_0^2=m_R^2 + B_\mathrm{div}$, and we arrive at ($\zeta=1+\delta\zeta$)
\begin{equation}
  \label{gapeq_fin}
  \zeta p^2 - m_R^2 = K(p) + \Sigma_\mathrm{fin}[K](p),
\end{equation}
which is a finite equation.

We will denote the kernel optimized by 2PI equations for calculation at
temperature $T$ as $K^{(T)}$. We also will denote a diagram or observable
$\Gamma$ computed with kernel $K$ and at temperature $T$ as $\Gamma[K,T]$. We
remark here that we can use a kernel, optimized to a certain temperature, in
computations at other temperatures as well. The optimization temperature is
just a notation to label the kernel, while there can be another real
temperature in the system. Weinberg theorem ensures that
$\Gamma^\mathrm{div}[K,
\begin{picture}(8,0)
  \put(0,0){\mbox{T}}
  \put(-2,-2){\line(1,1){10}}
\end{picture}],$ the overall divergent part does not depend on the (real)
temperature. 

The strategy to solve the finite temperature 2PI equations will consist of two
steps.
\begin{itemize}
\item[\textbf{Step 1.}] We determine the kernel $K^{(0)}$ which is optimized
  for zero temperature, ie.  the solution of \eqref{gapeq_fin} at $T=0$. The
  parameters $\zeta,\, m_R^2$ and $\lambda_R$ are determined by using physical
  observables. In the calculations below we just choose them to be $m_R=m$,
  $\zeta=1$ and $\lambda_R=\lambda$, so we should solve:
  \begin{equation}
    \label{1ststep}
    K^{(0)}(p)=  p^2 - m^2 - \Sigma_\mathrm{fin}[K^{(0)},0](p),
  \end{equation}
  We will use the normalization conditions for the self-energy as:
  \begin{equation}
    \label{renorm_at_T0}
    \Sigma_\mathrm{fin}[K^{(0)},0](p^2=m^2) = 0,\qquad 
    \frac{\d  \Sigma_\mathrm{fin}[K^{(0)},0]}{\d p^2} \biggr|_{p^2=m^2} =0.
  \end{equation}
  Then at $p^2=m^2$ we will find a pole of the propagator, with unit
  residuum.
\item[\textbf{Step 2.}] We fix the temperature at $T$, and solve
  \eqref{gapeq_fin} to have $K^{(T)}$. To be consistent with the $T=0$
  calculation we cannot just choose the renormalized parameters ( $\zeta_R,\,
  m_R$ and $\lambda_R$) of the Lagrangian, we will find their value by
  matching to the $T=0$ case. 

  There are two possible strategies here. The first is to find the solution by
  taking arbitrary finite part for $\Sigma$, but using undetermined values for
  $\zeta_R,\, m_R$ and $\lambda_R$. Then, using $K^{(T)}$ as a kernel, we
  compute the \emph{zero temperature} propagator and 4-point function. Since
  the kernel was not optimized for $T=0$ calculation, we do not expect
  especially good convergence of perturbative series, but we do expect that
  the asymptotic regime will be well described.  So we will choose large
  momenta (eg.  $p=(N m,0,0,0)$ where $N=20$ - $50$) and match
  $G^{-1}(p)\equiv K(p)$ and $\Gamma^{(4)}(p,0,-p,0)$ at these momenta
  (choosing together 3 matching conditions).
  
  The other possible strategy -- which will be followed later -- defines the
  finite part of $\Sigma$ in a way that it is possible to choose $\zeta_R=1$
  and $m_R=m$. Then the relevant equation here reads (cf. \eqref{gapeq_fin})
  \begin{equation}
    \label{2ndstep}
    K^{(T)}(p)=  p^2 - m^2 - \Sigma_\mathrm{fin}[K^{(T)},T](p).
  \end{equation}
  For matching we still use $G^{-1}(p)\equiv K(p)$ and
  $\Gamma^{(4)}(p,0,-p,0)$ at asymptotic momenta. The zero temperature
  propagator, calculated with the finite temperature kernel reads
  \begin{equation}
    G^{-1}[K^{(T)},0](p) = K^{(T)}(p) -\Sigma[K^{(T)},0](p) +\delta K_1(p).
  \end{equation}
  Since 2PI equation (ie. $\Sigma_\mathrm{tot}=0$) yields $\delta K_1(p) =
  \Sigma[K^{(T)},T](p)$, we obtain
  \begin{equation}
    G^{-1}[K^{(T)},0](p) = K^{(T)}(p) -\Sigma[K^{(T)},0](p) +
    \Sigma[K^{(T)},T](p). 
  \end{equation}
  The infinite part of $\Sigma[K^{(T)},0](p)$ and $\Sigma[K^{(T)},T](p)$ is
  the same (overall divergence is temperature independent), so we can write
  \begin{equation}
    G^{-1}[K^{(T)},0](p) = K^{(T)}(p) -\Sigma_\mathrm{fin}[K^{(T)},0](p) +
    \Sigma_\mathrm{fin}[K^{(T)},T](p).
  \end{equation}
  Now we use $K^{(T)}(p) +\Sigma_\mathrm{fin}[K^{(T)},T](p)= p^2-m^2$, then we
  find 
  \begin{equation}
    G^{-1}[K^{(T)},0](p) = p^2-m^2 - \Sigma_\mathrm{fin}[K^{(T)},0](p).
  \end{equation}
  This should be equal to
  \begin{equation}
    G^{-1}[K^{(0)},0](p)= K^{(0)}(p) = p^2 -m^2
    -\Sigma_\mathrm{fin}[K^{(0)},0](p).
  \end{equation}
  So the asymptotic equality of the two propagators requires that we should
  define the finite part of the self energy in a way that
  \begin{equation}
    \label{finpartT}
    \Sigma_\mathrm{fin}[K^{(0)},0](p) =  \Sigma_\mathrm{fin}[K^{(T)},0](p)
  \end{equation}
  shall be true for two asymptotic momenta.
\end{itemize}

After obtaining the optimal kernel, we can use ordinary perturbation theory to
compute other observables.

\section{$\Phi^4$ theory, two loop level, finite temperature}
\label{sec:Phi4}

Now let us test the above ideas in the case of the $\Phi^4$ theory at finite
temperature at two loop level. We will use real time formalism in R/A basis
\cite{RAformalism}. Here the original fields $(\Phi_1,\,\Phi_2)$ are replaced
by $(\Phi_r,\Phi_a)$ via the definition
\begin{equation}
  \Phi_1=\Phi_r +\frac12 \Phi_a,\qquad \Phi_2=\Phi_r -\frac12 \Phi_a.
\end{equation}
The propagator is a $2\times2$ matrix, where, in this basis $G_{aa}=0$
for the exact Green's function, and also $G^*_{ra}(p) = G_{ar}(p)$ is
true in the Fourier space. The interaction Lagrangian reads
\begin{equation}
  -{\cal L}_I =  \frac \lambda{24}\Phi_r^3\Phi_a + \frac\lambda{24}
  \Phi_r\Phi_a^3.
\end{equation}

A generic kernel is a $2\times2$ matrix, but it must respect symmetries valid
for any finite temperature field theory. In particular, all the propagators
should be derived from the spectral function:
\begin{equation}
  G_{11}=G_{ra}+G_{12},\quad G_{22}= -G_{ra}+G_{21},\quad iG_{12}(p) =n(p_0)
  \rh(p),\quad iG_{21}(p) = (1+n(p_0)) \rh(p),
\end{equation}
and
\begin{equation}
  \label{KK1}
  G_{ra}(p) = \int\limits_{-\infty}^\infty \!\frac{d\omega}{2\pi} \,
  \frac{\rh(\omega,\p)}{p_0-\omega+i\ep},\qquad \rh(p) =-2\Im G_{ra}(p).
\end{equation}

The retarded self-energy $G_{ra}$ can be expressed through its own self
energy, it will not mix with other components:
\begin{equation}
  G_{ra}^{-1}(p) = G_{ra,0}^{-1}(p) - \Sigma_{ar}(p).
\end{equation}
The other self energies form a matrix, and they can be obtained as
\begin{eqnarray}
  \label{sigmarels}
  && \Re \Sigma_{11} = -\Re\Sigma_{22} = \Re \Sigma_{ar},\qquad 
  \Im \Sigma_{11}= \Im \Sigma_{22}= (1+2n_B) \Im \Sigma_{ar},\nn&&
  \Sigma_{12}=-2in_B \Im \Sigma_{ar}, \qquad 
  \Sigma_{21}=-2i(1+n_B) \Im \Sigma_{ar}.
\end{eqnarray}
These relations are to be respected when we choose the generic kernel $K$. We
denote the retarded kernel by $K_{ra}$, the other matrix elements come from
\eqref{sigmarels}. The free retarded propagator then simply reads as
\begin{equation}
  G_{ra,0}^{-1}(p) = K_{ra}(p).
\end{equation}
The free spectral function can be expressed as
\begin{equation}
  \label{rho0}
  \rh_0(p)  = -2\Im  G_{ra,0}(p) =  \frac{-2 \Im
    K_{ra}(p)}{(\Re K_{ra}(p))^2 +(\Im K_{ra}(p))^2}.
\end{equation}
Causality requires that $\Im K_{ra}(p)$ must be an odd function of $p_0$, and
also $\Im K_{ra}(p)< 0$ must be true for $p_0>0$. In the \MSbar\ scheme it is
satisfied by choosing $\Im K_{ra}(p)=-\ep\, \sgn p_0$ (Landau prescription).

\subsection{One loop level}

At one loop we do not generate momentum dependence for the kernel. This
problem has been discussed with the momentum-independent version of the
present method in \cite{JakovacSzep}, but we include it here for completeness
and for elucidate the strategy. The self-energy and the one-loop four point
function read:
\begin{equation}
  \Sigma^{(1)}[K](p) = \frac\lambda2 \c T[K],\qquad
  \Gamma^{(4)}(p,q,k,\ell) = \lambda +\frac{\lambda^2}2 \biggl\{ I[K](p+q) +
  I[K](p+k) +I[K](p+\ell) \biggr\} +\delta\lambda_1.
\end{equation}
where $\c T[K]$ is the tadpole, $I[K]$ is the bubble diagram, computed with
kernel $K$. The tadpole can be expressed as $G_{11}(x=0) = G_{12}(x=0)$; in
Fourier space
\begin{equation}
  \c T[K]= \pint4k n(k_0)\rh(k) = \int\limits_{k_0>0}\frac{d^4k}{(2\pi)^4} (1
  + 2n(k_0)) \rh(k).
\end{equation}
To compute the bubble diagram one should consult Appendix \ref{sec:bubble}.
For the 2PI summation we choose
\begin{equation}
  \label{deltaK1}
  \delta K_1=\frac\lambda2 \c T[K].
\end{equation}

For \textbf{Step 1} we have to define its finite part so that the self-energy
shall be zero on the mass shell at zero temperature. It results in the
definition
\begin{equation}
  \label{tadpolesplit}
  \c T[K]= \c T_\mathrm{div}[K]+\c T_\mathrm{fin}[K],\qquad \c
  T_\mathrm{div}[K] = \int\limits_{k_0>0}\frac{d^4k}{(2\pi)^4} \rh(k),\qquad
  \c T_\mathrm{fin}[K] = \int\limits_{k_0>0}\frac{d^4k}{(2\pi)^4}\, 2n(k_0)
  \rh(k).
\end{equation}
With this definition the zero temperature equation to solve is
(cf. \eqref{1ststep})
\begin{equation}
  K^{(0)}(p) = p^2 -m^2,
\end{equation}
since the self energy is zero at zero temperature.

For \textbf{Step 2}, with proper finite part for $\Sigma$, we have to solve
\eqref{2ndstep}:
\begin{equation}
  \label{oneloopfinT}
  K^{(T)}(p) =  p^2 - m^2 - \frac{\lambda_T}2 \c T_\mathrm{fin}[K^{(T)},T]. 
\end{equation}
In order to be write this equation, we have to ensure that the self energies
at zero temperature, calculated with kernels $K^{(0)}$ and $K^{(T)}$ are
equal. But now the zero temperature self energy, according to
\eqref{tadpolesplit}, is simply zero, so this requirement is trivially
satisfied. 

\eqref{oneloopfinT} is the same result that was obtained also in
\cite{JakovacSzep}.

For the \textbf{coupling constant renormalization} we choose for kernel $K$ the
counterterm
\begin{equation}
  \label{oneloopdlambda}
  \delta \lambda_1 = -\frac{3\lambda^2}2 I[K,0](p=0),
\end{equation}
and define the finite part of the bubble diagram as
\begin{equation}
  I_\mathrm{fin}[K,T](p,q,k,\ell) =   I[K,T](p,q,k,\ell) - I[K,0](0).
\end{equation}
This is of course just a specific choice, we could add a finite term to it in
another renormalization scheme. The above choice ensures that the renormalized
one loop four point function at zero momentum at zero temperature is
$\lambda_R=\lambda$.

We use this equation to write the second matching condition as
\begin{eqnarray}
  \label{lambda2matching}
  &&\lambda +\frac{\lambda^2}2 \biggl[ I_\mathrm{fin}[K^{(0)},0](p+q) +
  I_\mathrm{fin}[K^{(0)},0](p+k) +I_\mathrm{fin}[K^{(0)},0](p+\ell)
  \biggr] =\nn&&=  \lambda_T +\frac{\lambda_T^2}2 \biggl[
  I_\mathrm{fin}[K^{(T)},0](p+q) + I_\mathrm{fin}[K^{(T)},0](p+k)
  +I_\mathrm{fin}[K^{(T)},0](p+\ell) \biggr],
\end{eqnarray}
at asymptotic momenta. In this simple case only the mass is modified, let us
denote the mass of the kernel $K$ as $M$. Then we can explicitly write up
these contributions:
\begin{eqnarray}
  && I[K,0](0) = \frac1{16\pi^2}\left[-\frac1\ep +\gamma_E
    +\ln\frac{M^2}{4\pi\mu^2} \right]\nn
  && I_\mathrm{fin}[K,0](p^2>4M^2) =  \frac1{16\pi^2}\left[
    \ln\left|\frac{p^2}{M^2}\right| + 2X_+ (\ln X_+ -1)  + 2X_- (\ln X_- -1)
  \right],
\end{eqnarray}
where
\begin{equation}
  X_\pm = \frac12 \left(1\pm \sqrt{1-\frac{4M^2}{p^2}}\right).
\end{equation}
For asymptotically large momenta ($p^2\gg M^2$) we have
\begin{equation}
  I_\mathrm{fin}[K,T=0](p) = \frac1{16\pi^2} \ln p^2 + \c O(\mathrm{const.})
\end{equation}
Therefore, in leading order in momenta, equation \eqref{lambda2matching} can
be written (choosing $p=q=-k=-\ell$) as
\begin{equation}
  \lambda +\frac{\lambda^2}{32\pi^2} \ln p^2 = \lambda_R
  +\frac{\lambda_R^2}{32\pi^2} \ln p^2,
\end{equation}
which implies $\lambda_T=\lambda$.

We will use this matching condition also at two loop level for simplicity. But
we can also argue that -- since the asymptotic parts of the kernels $K^{(0)}$
and $K^{(T)}$ are the same (cf. \eqref{finpartT}) -- we obtain the same
$k$-dependence for the 4-point function at asymptotic momenta using either
$K^{(T)}$ or $K^{(0)}$. Then $\lambda_R=\lambda$ should be the result at all
order.

\subsection{Two loop analysis}

Let us repeat the above analysis at two loop level. From the generic strategy
we assume that the kernel is fixed throughout the calculation. The diagrams
for the self energy read:
\begin{equation}
  \label{twoloopSigma}
  \Sigma^{(2)}[K](p) = \frac{\lambda^2}6
  S[K](p) + \frac{\lambda^2}4 \c T[K]\, I[K](0) - \frac{\lambda\delta K_1}2
  I[K](0) + \frac{\delta\lambda_1}2 \c T[K], 
\end{equation}
where $\c T[K]$ stands for tadpole, $I[K]$ for bubble, $S[K]$ for setting
sun diagrams. The computation of the setting sun diagram is discussed in
Appendix \ref{sec:setsun}. $\delta K_1$ and $\delta \lambda_1$ is fixed at
one-loop level (cf. \eqref{deltaK1} and \eqref{oneloopdlambda}). 

To see the overall divergence we split the setting sun diagram as $S[K](p) =
S_\mathrm{fin}[K](p) + S_\mathrm{subdiv}[K]+ S_\mathrm{overall}[K](p),$ where
\begin{equation}
  \label{setsunsep}
  S_\mathrm{subdiv}[K] = 3\c T[K] \c I[K,0](0) = -2\delta\lambda_1 \c
  T[K],\qquad S_\mathrm{overall}[K](p)= \delta Z_S p^2 -\delta m_S^2,
\end{equation}
and $\delta Z_S$ and $\delta m_S$ depend on the kernel and the choice of the
renormalization conditions. We find
\begin{equation}
  \Sigma^{(2)}[K](p) = \frac{\lambda^2}6 S_\mathrm{fin}[K](p) + \delta Z_S p^2
  -\delta m_S^2+ \frac{\delta\lambda_1}6 \c T[K],
\end{equation}
We want to make the total self-energy correction disappear in the spirit of
the 2PI resummation (cf. \eqref{2PIeq}), so we should choose $\delta K_2(p)
=\Sigma^{(2)}[K](p)$. To tell the bare parameters of the Lagrangian we should
write (cf. eq.  \eqref{bare_ren_relation})
\begin{equation}
  \label{gapeq2}
  Zp^2 - m_0^2 = K(p) +  \delta K_1 +  \delta K_2.
\end{equation}
We write $m_0^2 = m_R^2 + \delta m_1^2 +\delta m_2^2$ and $Z=1+\delta
Z_1+\delta Z_2$. Then we identify
\begin{equation}
  \delta m_1^2=-\delta K_1= -\frac\lambda2 \c T_\mathrm{div},\qquad \delta
  m_2^2 =  \delta m_S^2 - \frac{\delta\lambda_1}6 \c T[K],\qquad \delta
  Z_1=0,\qquad \delta Z_2=-\delta Z_S.
\end{equation}

To renormalize the theory we should perform the two steps described earlier:

\textbf{Step 1} we have to solve the zero temperature 2PI equation
\eqref{1ststep}:
\begin{equation}
  K^{(0)}(p) = p^2 -m^2  - \frac{\lambda^2}6 S_\mathrm{fin}[K^{(0)},0](p),
\end{equation}
since $\c T_\mathrm{fin}[K^{(0)},0]=0$. We should define the finite part as 
\begin{equation}
  \label{deffinsetsun1}
  S_\mathrm{fin}[K^{(0)},0](p^2=m^2) = 0,\qquad 
  \frac{\d  S_\mathrm{fin}[K^{(0)},0]}{\d p^2} \biggr|_{p^2=m^2} =0.
\end{equation}
Then the above equation yields a propagator which has a pole at $p^2=m^2$ with
unit residuum.

In numerical calculations one has to compute the complete contribution,
subtract $3\c T[K^{(0)},0] \c I[K^{(0)},0](0)$ subdivergence, and subtract an
$ap^2-b$ function.  The value of $a$ and $b$ can be determined from eq.
\eqref{deffinsetsun1}.

\textbf{Step 2} the finite temperature equation to solve is \eqref{2ndstep}:
\begin{equation}
  K^{(T)}(p) = p^2 -m^2 -\frac\lambda2\c T_\mathrm{fin}[K^{(T)},T] -
  \frac{\lambda^2}6 S_\mathrm{fin}[K^{(T)},T](p).
\end{equation}
For the correct renormalization we should require \eqref{finpartT}:
\begin{equation}
  \label{deffinsetsun2}
  S_\mathrm{fin}[K^{(T)},0](p) = S_\mathrm{fin}[K^{(0)},0](p) \qquad
  \mathrm{for\ asymptotic\ momenta}.
\end{equation}

Numerically we should compute the complete setting sun contribution with
kernel $K^{(T)}$ at zero temperature, subtract $3\c T[K^{(T)},0] \c
I[K^{(T)},0](0)$ subdivergence, and finally $Ap^2-B$. We determine $A$ and $B$
from \eqref{deffinsetsun2}, by requiring its fulfillment at two asymptotic
momenta $p_1$ and $p_2$. Then at finite temperature we should compute the
renormalized value of the setting sun diagram, that we first compute the
complete diagram with kernel $K^{(T)}$ at finite temperature, subtract $3\c
T[K^{(T)},T] \c I[K^{(T)},0](0)$ subdivergence, and also $Ap^2-B$ with the
values determined before. This will cancel all divergences, since, as it was
mentioned earlier, the overall divergences do not depend on the temperature.

\subsection{Results}

The above strategy can be accomplished by numerical calculations. We used the
spatial rotational invariance of the propagators, which allows to represent
them at finite temperature as two-dimensional functions, with variables $p_0$
and $|\p|$. Therefore for the integrations we need a 2D lattice, which is
mapped with a continuous function to the infinite 2D space. All the results
presented here are calculated at $\lambda=10$. Finite temperature, if not
stated otherwise, means here $T=m$.

The 2PI equations are solved by successive approximation, starting from a
spectral function near to the free one. From the spectral function one can
determine the propagators, with the propagators we calculate the
self-energies, and with help of self energies we update the spectral function.
Then repeat this algorithm till it converges.

\begin{figure}[htbp]
  \centering
  \includegraphics[height=5cm]{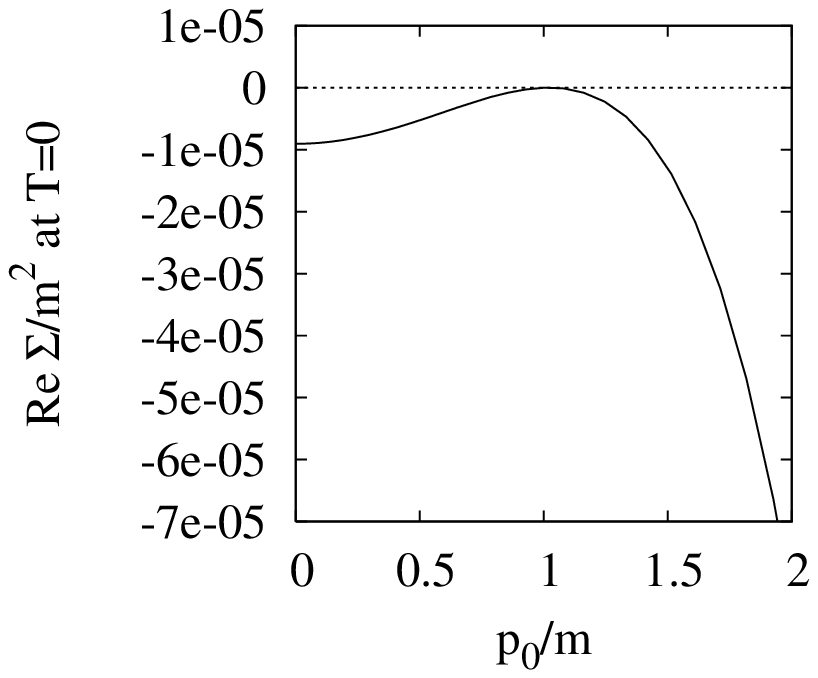}
  \includegraphics[height=5cm]{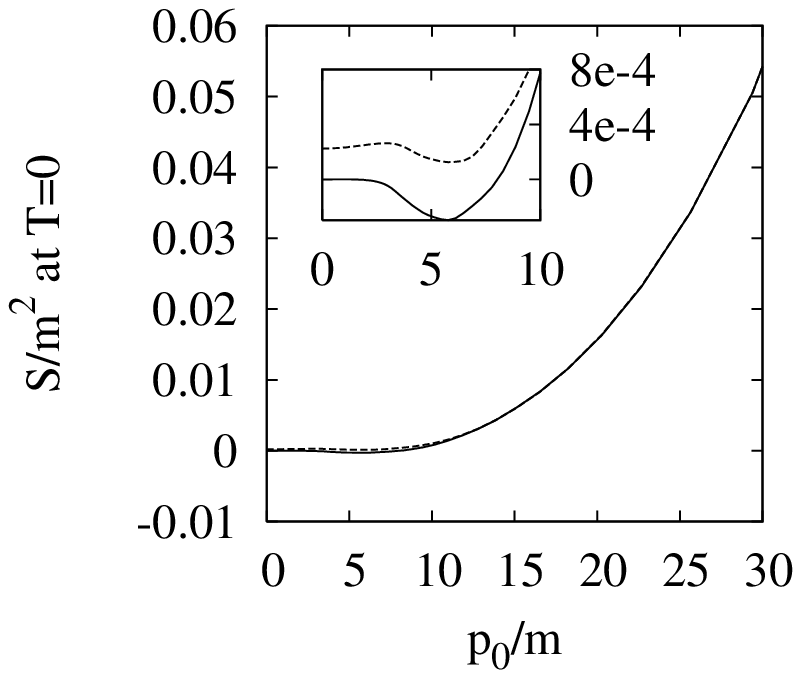}\\
  \hspace*{0.5cm}a.)\hspace*{6.8cm}b.)
  \caption{Quality of renormalization: a.) zero temperature self-energy at
    $\mathbf{p}=0$; b.) zero temperature setting sun diagram with different
    kernels -- the two lines are almost within line width.}
  \label{fig:renorm}
\end{figure}

First we discuss the renormalization. In Fig. \ref{fig:renorm}
we can see the figures representing the renormalization process. On Fig.
\ref{fig:renorm}/a there appears the self energy calculated with $K^{(0)}$. It
can clearly be seen that at $p_0=m$ the self energy is $\Sigma=0$, so there is
a pole there. It can also be seen that the derivative with respect to $p_0$ is
zero -- this ensures the unit residuum. Fig. \ref{fig:renorm}/b concerns the
finite temperature renormalization. Here we require that the values of setting
sun diagram calculated with kernel $K^{(0)}$ or $K^{(T)}$ at zero temperature,
are, asymptotically, the same. As this plot shows, this is fulfilled within
line width, although for smaller momenta there can be seen some discrepancy.
These two plots show that the theory is renormalized.

\begin{figure}[htbp]
  \centering
  \includegraphics[height=5cm]{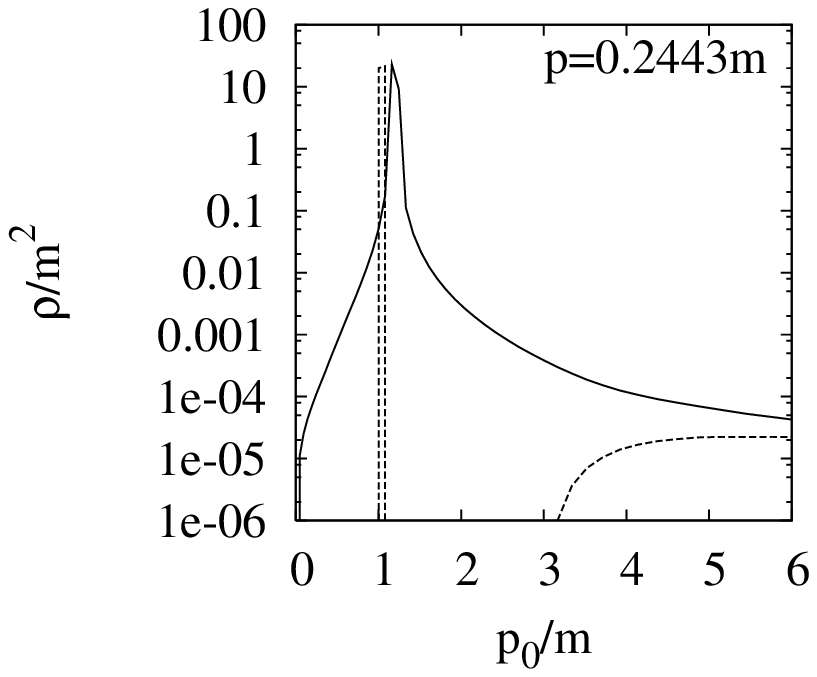}
  \includegraphics[height=5cm]{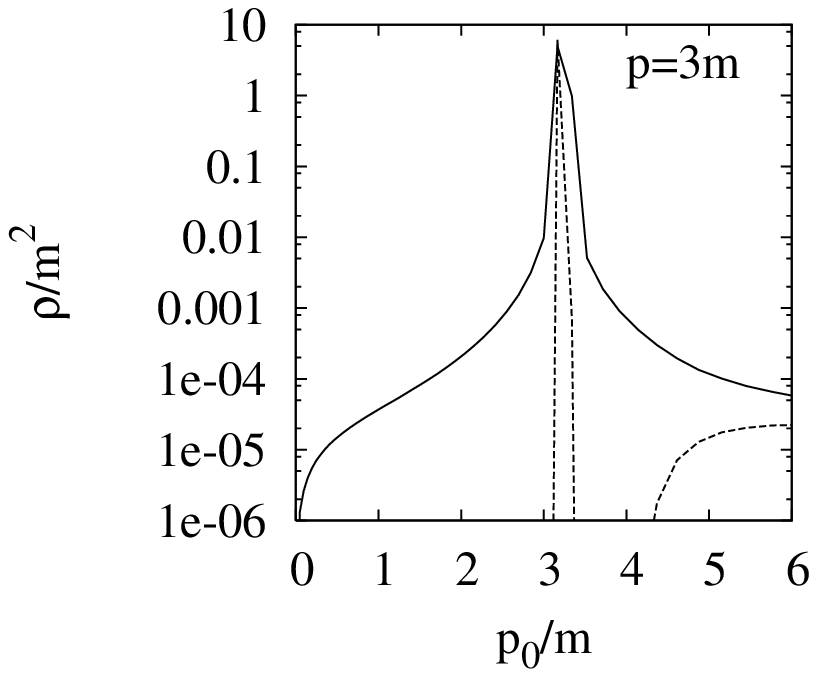}\\
  \hspace*{0.8cm}a.)\hspace*{7cm}b.)
  \caption{Spectral function at two values of the spatial momentum at zero
    temperature (dashed line) and at finite ($T=m$) temperature (solid
    line). The $y$ axis is logarithmic.}
  \label{fig:spectral}
\end{figure}

The next two plots (Fig. \ref{fig:spectral}) show the spectral function at two
different spatial momentum.
At zero temperature one can clearly see the quasiparticle peak, and a well
separated threshold starting at $3m$ in case of $\mathbf{p}=0$. The position
of the peak as well as the threshold value shifts with the spatial momentum.
At finite temperature the quasiparticle peak is still very sharp and narrow
(at these values of the coupling), but there is no sign for any threshold
behavior, only a broad continuum can be observed.

\begin{figure}[htbp]
  \centering
  \includegraphics[height=5cm]{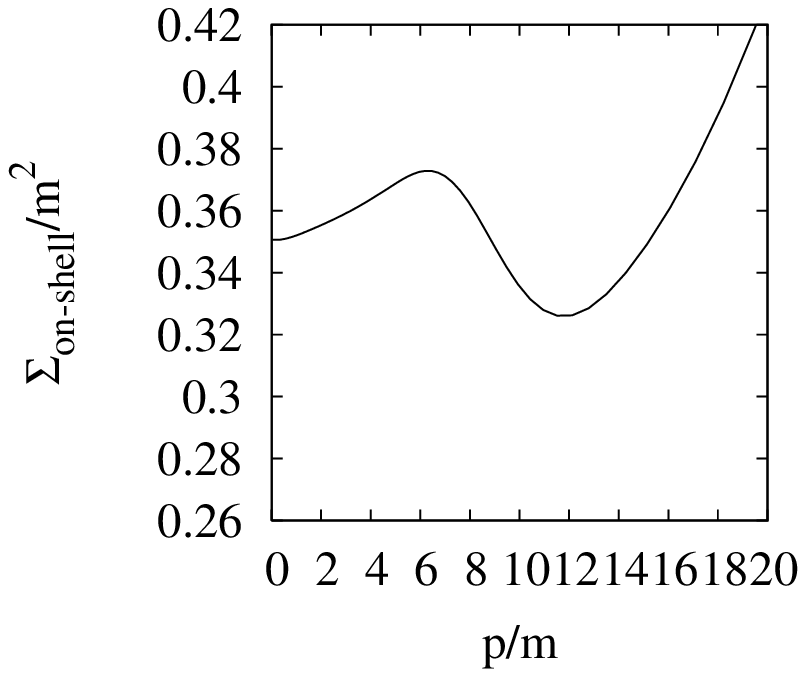}
  \includegraphics[height=5cm]{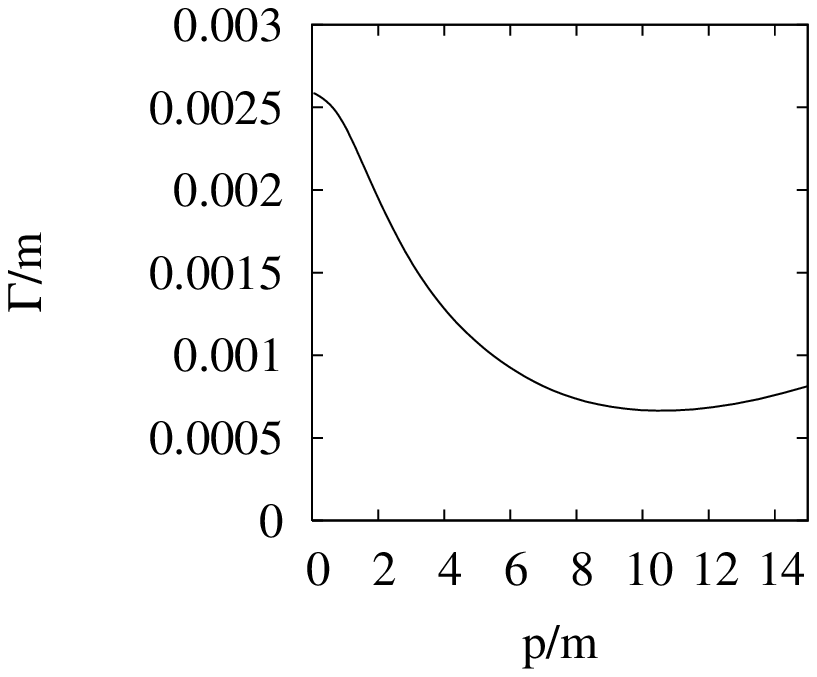}\\
  \hspace*{0.9cm}a.)\hspace*{7.3cm}b.)
  \caption{The quasiparticle properties: spatial momentum dependence of the 
    mass (a.) and the width (b.).}
  \label{fig:quasi}
\end{figure}

In the third pair of figures (Fig. \ref{fig:quasi}) the quasiparticle
properties can be seen.
The first plot shows the spatial momentum dependence of the mass. The momentum
dependence is rather weak: in the range $0<|\mathbf{p}|<15m$ it is less than
10\%. This fact, together with the dominant quasiparticle peak may explain the
success of the ``screened perturbation theory'' \cite{screenedpert}. In the
second plot the quasiparticle width can be seen: it drops\footnote{The
  apparent rise of the width is lattice effect: we used variable lattice
  spacing, growing at larger momenta.} about a factor of 4 in the range
$0<|\mathbf{p}|<10m$, as was predicted by perturbative calculations
\cite{HeinzWang}.

The last two plots demonstrate the temperature dependence of the quasiparticle
properties.
\begin{figure}[htbp]
  \centering
  \includegraphics[height=5cm]{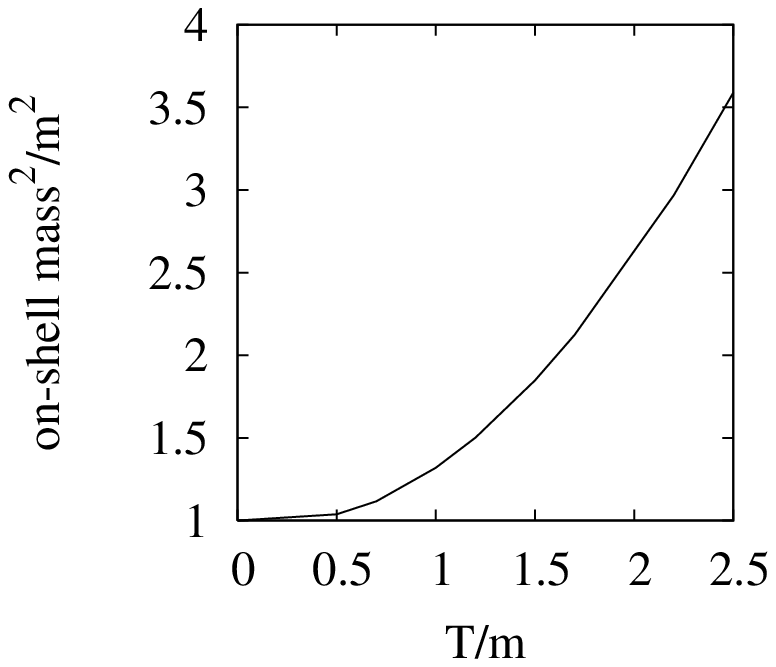}
  \includegraphics[height=5cm]{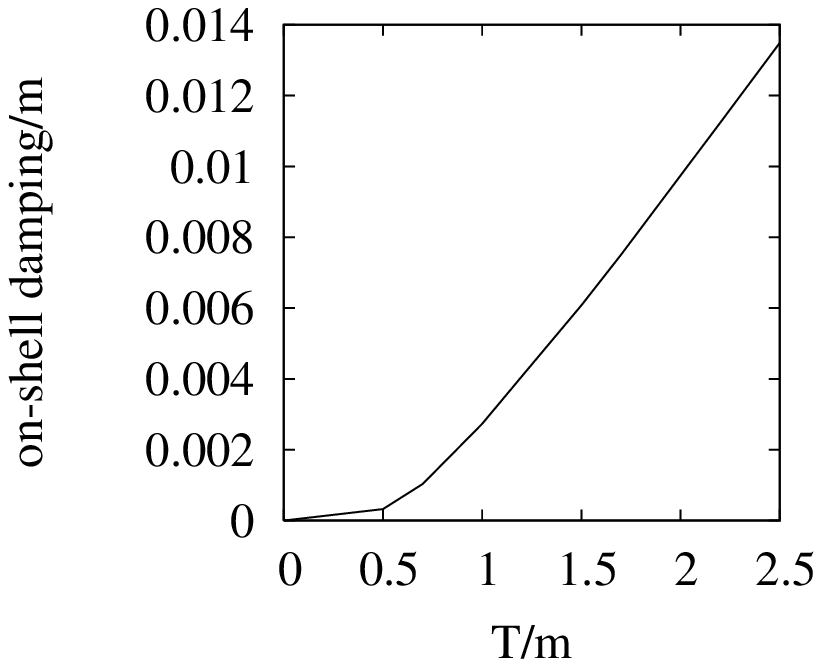}\\
  \hspace*{0.5cm}a.)\hspace*{7.3cm}b.)
  \caption{The temperature dependence of the quasiparticle properties: mass
    (a.) and the width (b.).}
  \label{fig:tempdep}
\end{figure}
In Fig. \ref{fig:tempdep}/a. the temperature dependence of the on-shell mass
can be seen. In the range $T<0.5m$ there is hardly any temperature dependence;
at $T>1.2m$ the curve fits well to a second order polynomial, so there
$m^2=m_0^2 + cT^2$ is a good assumption. Fig.  \ref{fig:tempdep}/b shows the
on-shell damping rate as a function of the temperature. The same tendency can
be observed as in the on-shell mass case: for small temperatures ($T<0.5m$)
there is hardly any damping can be seen, while for high temperatures ($T>m$)
the damping rate can be fitted to a straight line $\Gamma=\Gamma_0+cT$. In
both cases the $T\approx m$ regime is a crossover between the two
characteristic behavior.

\section{Conclusion}
\label{sec:conclusion}

The most important message of this paper is that the ``renormalization scheme
method'' is appropriate to perform fully renormalized momentum dependent 2PI
resummation at finite temperature. The advantage of this method is that one
does not need to solve the Bethe-Salpeter equations. Instead one fixes the
renormalization in the finite temperature calculation by matching to the zero
temperature results.

Renormalization scheme method is based on the observation that in certain
momentum dependent schemes there is no need of doing 2PI resummation, since,
by construction, the self energy is zero at each order. We studied in the
paper, when such an ``optimized'' scheme yields a consistent perturbation
theory. We examined the relation of two such schemes, and proved that by
appropriate choice of the renormalized parameters, two schemes can give the
same result up to $\c O(\lambda^n)$, where $n$ is the order of the
perturbation theory -- this is the same result as in the ``ordinary'' momentum
independent schemes. But it also suggests that if we fix a scheme at zero
temperature, then at finite temperature the optimized scheme can be related to
it by matching.

In particular we suggested in the paper a two-step method for the
renormalization. In the first step one has to solve the zero temperature 2PI
scheme equations, where we fix its parameters by observable quantities (like
quasiparticle mass, residuum and scattering cross section). In the second step
we solve the finite temperature 2PI scheme equations, defining the finite part
of the self energy in a specific way: we require that the zero temperature
self energy, calculated with the zero temperature and the finite temperature
optimized kernel, are the same for asymptotic momenta. This definition is the
manifestation of the matching procedure.

The so-defined rules are numerically easily implementable, as it was
demonstrated in the paper on the example of the $\Phi^4$ theory with two-loop
2PI resummation. The results of this study are summarized in the paper. The
most important corollary is that the $\Phi^4$ theory can be well characterized
by a thermal mass: the spectral function is dominated by a quasiparticle peak
where its spatial momentum dependence can be described by an almost constant
mass term.

For future studies, we plan to generalize the renormalization scheme method to
non-equilibrium case, and also -- which may be even more important and
interesting -- to systems with specific symmetries (global or gauge).

\section*{Acknowledgment}

The author acknowledges useful discussions with T. S. B\'{\i}r\'o, A.
Patk\'os, and Zs. Sz\'ep. This work was supported by the Hungarian grant OTKA
F043465.

\appendix

\section{Diagrams}

\subsection{Bubble}
\label{sec:bubble}

We compute the retarded diagram that reads
\begin{equation}
  iI[K,T](k) = \pint4p \left(iG_{11}(p) iG_{11}(k-p) - iG_{12}(p)
  iG_{12}(k-p) \right).
\end{equation}
Using the dispersion relation
\begin{equation}
  G_{ra}(k) = \pintz s\,\frac{\rh(s,\k)}{k_0-s+i\ep}
\end{equation}
the retarded diagram can be computed from its discontinuity
\begin{eqnarray}
  && \Disc_{k_0} iI[K,T](k) = \pint4p (1+ n(p_0) +n(k_0-p_0)) \rh_1(p)
  \rh_2(k-p),\nn
  && I[K,T](k) = \pintz s\,\frac{\Disc_s iI[K,T](s,\k)} {k_0-s+i\ep}.
\end{eqnarray}
At zero temperature
\begin{equation}
  n(\omega) = \frac1{e^{\beta\omega}-1}
  \stackrel{\beta\to\infty}{\longrightarrow} -\Theta(-\omega).
\end{equation}
Thus
\begin{equation}
   1+ n(p_0) +n(k_0-p_0) = \Theta(p_0) - \Theta(p_0-k_0) = \Theta(0<p_0<k_0).
\end{equation}
So
\begin{equation}
  \Disc_{k_0}iI[K,0](k) =  \pint4p \Theta(0<p_0<k_0) \rh_1(p)\rh_2(k-p).
\end{equation}

\subsection{Setting sun diagram}
\label{sec:setsun}

The setting sun diagram can be reproduced from its discontinuity, just like
the bubble. The discontinuity reads
\begin{eqnarray}
  \Disc iS[K](k) &&= \pint4p\frac{d^4q}{(2\pi)^4} \left[iG_{21}(p)\,iG_{21}(q)
    \, iG_{21}(\ell) - iG_{12}(p)\,iG_{12}(q) \,iG_{12}(\ell)\right] \nn&&=
  \pint4p\frac{d^4q}{(2\pi)^4}\,\left[ (1+ n_p)(1+ n_q)(1+n_\ell) - n_p n_q
    n_\ell\right] \rh_1(p) \rh_2(q) \rh_3(\ell), 
\end{eqnarray}
where $p+q+\ell=k$. At zero temperatures it falls back to
\begin{equation}
  \Disc iS[K,T=0](k) = \pint4p\frac{d^4q}{(2\pi)^4}\, \rh_1(p) \rh_2(q)
  \rh_3(\ell) \left[\Theta(p_0)\Theta(q_0)\Theta(\ell_0) -
    \Theta(-p_0)\Theta(-q_0)\Theta(-\ell_0)\right].
\end{equation}


\begin{thebibliography}{99}
\bibitem{Collins} J. Collins, \emph{Renormalization} (Cambridge
  University Press, 1984).
\bibitem{HeesKnoll} H. van Hees an J. Knoll, Phys. Rev. D65 (2002) 025010
  [hep-ph/0107200]; Phys. Rev. D65 (2002) 105005 [hep-ph/0111193]; Phys. Rev.
  D66 (2002) 025028 [hep-ph/0203008]
\bibitem{Blaizotetal} J.-P. Blaizot, E. Iancu and U. Reinosa, Phys. Lett. B568
  (2003) 160 [hep-ph/0301201];
% Renormalizability of Phi derivable approximations in scalar phi**4
% theory.
  Nucl. Phys. A736 (2004) 149 [hep-ph/0312085]
%Renormalization of Phi derivable approximations in scalar field
%theories.
\bibitem{Bergesetal} J. Berges, Sz. Borsanyi , U. Reinosa and J.
  Serreau, Annals Phys. 320 (2005) 344 [hep-ph/0503240]
% Nonperturbative renormalization for 2PI effective action techniques
\bibitem{2PIgauge} U. Reinosa, J. Serreau, JHEP 0607 (2006) 028
  [hep-th/0605023]; %gauge theory 
\bibitem{Verschelde} H. Verschelde, Phys. Lett. B 497 (2001) 165
  [hep-th/0009123] 
\bibitem{PSz} A. Patkos, Zs. Szep, Phys. Lett. B642 (2006) 384
  [hep-th/0607143]
\bibitem{Cooperetal} F. Cooper, B. Mihaila and J. F. Dawson, Phys. Rev. D70
  (2004) 105008 [hep-ph/0407119]
% Renormalizing the Schwinger-Dyson equations in the auxiliary field
% formulation of lambda phi**4 field theory: 
\bibitem{Destrietal} C. Destri and A. Sartirana, Phys. Rev. D72 (2005) 065003
  [hep-ph/0504029]; Phys. Rev. D73 (2006) 025012 [hep-ph/0509032]
% The Renormalized and renormalization-group invariant Hartree-Fock
% approximation. 
\bibitem{ON1perN} J. O. Andersen, D. Boer and H. J. Warringa, Phys. Rev. D70
  (2004) 116007 [hep-ph/0408033]
\bibitem{JakovacSzep} A. Jakovac and Zs. Szep, Phys. Rev. D 71 (2005) 105001
  [hep-ph/0405226]
\bibitem{envfriend} D. O'Connor and C.R. Stephens, Int. J. Mod. Phys. A 9
  (1994) 2805; Erratum-ibid. A 9 (1994) 5851.
\bibitem{AJ} A. Jakovac, Phys. Rev. D74 (2006) 085026 [hep-ph/0605071]
\bibitem{Weinberg_thm} S. Weinberg, Phys. Rev. 78 (1960) 182;
  Y. Hahn and W. Zimmermann, Comm. Math. Phys. 10 (1968) 330.
\bibitem{AppelquistCarrazone} T. Appelquist and J. Carazzone, Phys. Rev. D11
  (1975) 2856
\bibitem{dimred} K. Kajantie, M. Laine, K. Rummukainen and M. Shaposhnikov,
  Nucl. Phys. B458 (1996) 90 [hep-ph/9508379]
\bibitem{CJT} J.M. Cornwall, R. Jackiw and E. Tomboulis, Phys. Rev. D10 (1974)
  2428
\bibitem{RAformalism} E. Wang and U. Heinz, Phys. Lett. B471 (1999) 208;
  K.-C.  Chou, Z.-B. Su, B.-L. Lao and L. Yu, Phys. Rep. 118 (1985) 1
\bibitem{screenedpert} F. Karsch, A. Patk\'os, P. Petreczky, Phys. Lett. B401
  (1997) 69 [hep-ph/9702376]
\bibitem{HeinzWang} E. Wang and U. Heinz, Phys. Rev. D53 (1996) 899
  [hep-ph/9509333]
\end{thebibliography}
\end{document}